\documentclass[%
reprint,
superscriptaddress,
amsmath,amssymb,
aps,
pra,
]{revtex4-2}

\usepackage{dcolumn}
\usepackage{bm}
\usepackage{float}
\usepackage{xcolor}
\usepackage{ulem}
\usepackage{graphicx,subfig}
\usepackage{adjustbox}
\usepackage{placeins}
\usepackage{hyperref}

\begin{document}
	
	\preprint{APS/123-QED}
	
	\title{Temperature control of diffusive memristor \\ hysteresis and artificial neuron spiking}
	\author{D.P Pattnaik}
	\email[]{D.Pattnaik@lboro.ac.uk}
	\affiliation{Department of Physics, Loughborough University, Loughborough LE11 3TU, United Kingdom}
	
	\author{Y. Ushakov}
	\thanks{DP and YU contributed equally to the paper}
	\affiliation{Department of Physics, Loughborough University, Loughborough LE11 3TU, United Kingdom}
	\author{Z. Zhou}
	\affiliation{Loughborough Materials Characterisation Centre, Loughborough University, Loughborough LE11 3TU, United Kingdom}
	
	\author{P. Borisov}
	\affiliation{Department of Physics, Loughborough University, Loughborough LE11 3TU, United Kingdom}
	
	\author{M.D. Cropper}
	\affiliation{Department of Physics, Loughborough University, Loughborough LE11 3TU, United Kingdom}
	\author{U.W. Wijayantha}
	\affiliation{Department of Chemistry, Loughborough University, Loughborough LE11 3TU, United Kingdom}
	\author{A.G. Balanov}
	\affiliation{Department of Physics, Loughborough University, Loughborough LE11 3TU, United Kingdom}
	\author{S. E. Savel’ev}
	\email[]{S.Saveliev@lboro.ac.uk}
	\affiliation{Department of Physics, Loughborough University, Loughborough LE11 3TU, United Kingdom}
	

	\begin{abstract}
	Memristive devices are promising elements for energy-efficient neuromorphic computing and future artificial intelligence systems. For diffusive memristors, the device state switching occurs because of the sequential formation and disappearance of conduction pillars between device terminals due to drift and diffusion of Ag nanoparticles in the dielectric matrix.
	This process is governed by application of the voltage to the device contacts. Here, both in experiment and in theory we demonstrate that varying temperature offers an efficient control of memristor states and charges transport in the device. We found out that by raising and lowering the device temperature, one can reset the memristor state as well as change the residual time the memristor stays in high and low resistive states when the current spiking is generated in the memristive circuit at a constant applied voltage. Our theoretical model demonstrates a good qualitative agreement with the experiments, and helps to explain the effects reported.
	\end{abstract}
\maketitle
\section{\label{sec:level1}Introduction}
In its simplest form, a memristor consists of an insulating or semiconducting film
sandwiched between two metallic electrodes. The switching layer~\cite{zhu2020comprehensive,lee2018demand}  undergoes dynamic material reconfiguration
under the influence of external electric field or current. This reconfiguration manifests as a recurring change in the device resistance switching between a low resistance state (LRS) and a high resistance state (HRS) observed in various devices and films~\cite{sawa2008resistive,waser2009redox,hickmott1962low,liu2020two}.

Among different types of memristors, diffusive memristors which are based on metal atom diffusion and spontaneous formation and rupture of conducting filament (CF)
hold much promise for the next generation neuromorphic
applications~\cite{jeong2012emerging,yang2013memristive,sangwan2020neuromorphic}.
The general mechanism of this switching process is explained by the electric field induced migration of metal atoms, in the form of metal nanoparticles (NPs) suspended in a dielectric
matrix~\cite{strukov2009exponential,sun2014direct,wang2017memristors}. Such a migration
can be modeled by an electrically biased diffusion, where the diffusion constant changes
in time due to Joule heating~\cite{savel2013mesoscopic} affecting device temperature and, thus, NP diffusion constant.  The external field promotes the metal nanoparticles to form 
CFs between the electrodes. The formation and dissolution of these CFs is dictated by the surface energy of the
nanoparticles.~\cite{hsiung2010formation,yang2012observation}

Due to inhomogeneities in the composition of the dielectric film~\cite{gaba2013stochastic} and roughness of the electrodes, the {NPs forming the CFs can be pinned~\cite{liu2010controllable,onofrio2015atomic}}, this affects CF rupture and the endurance and eventually can cause degradation of the memristive devices. In this case, degradation of a memristor is manifested as a device failure in which the memristor resistance is stuck in  LRS due to a permanent CFs formed between the electrodes~\cite{lv2015evolution}. Subsequent voltage sweeps can not rupture the CFs and reset the memristor resistance state. This is a major drawback of these diffusive memristor devices for technological applications where endurance and retention time are key characteristics~\cite{lanza2021standards,
baek2004highly,banerjee2020challenges}.
To date, no approaches have been proposed to resolve this limitation or possible physical methods to reset the conduction filament. 
Together with the bias voltage, the  Joule heating plays an important role in nanoparticle dynamics, however the effects of the dielectric matrix temperature on the process of filament formation and the related charge transport are poorly understood. In this work, we experimentally studied how temperature of the diffusive memristor influences its current-voltage characteristics and voltage oscillations generation in a memristive circuit  of an artificial neuron. Our theoretical analysis confirmed  that the device temperature can regulate and reconfigure the formation and dissolution of the CFs in diffusive memristors comprising Ag nanoparticles  suspended in a SiO$_x$ dielectric matrix.  We found out that by appropriate tuning of the temperature one can not only reset the resistive memristor states, but also  efficiently control the dynamical regimes of a memristive circuit.  Our theoretical modeling agrees well  with  the results of experiments. Thus, our findings open another avenue for development  and design of memristive elements for future electronics.
\begin{figure*}[ht]
	\centering
	\subfloat[] {
		\includegraphics[width=0.42\textwidth]{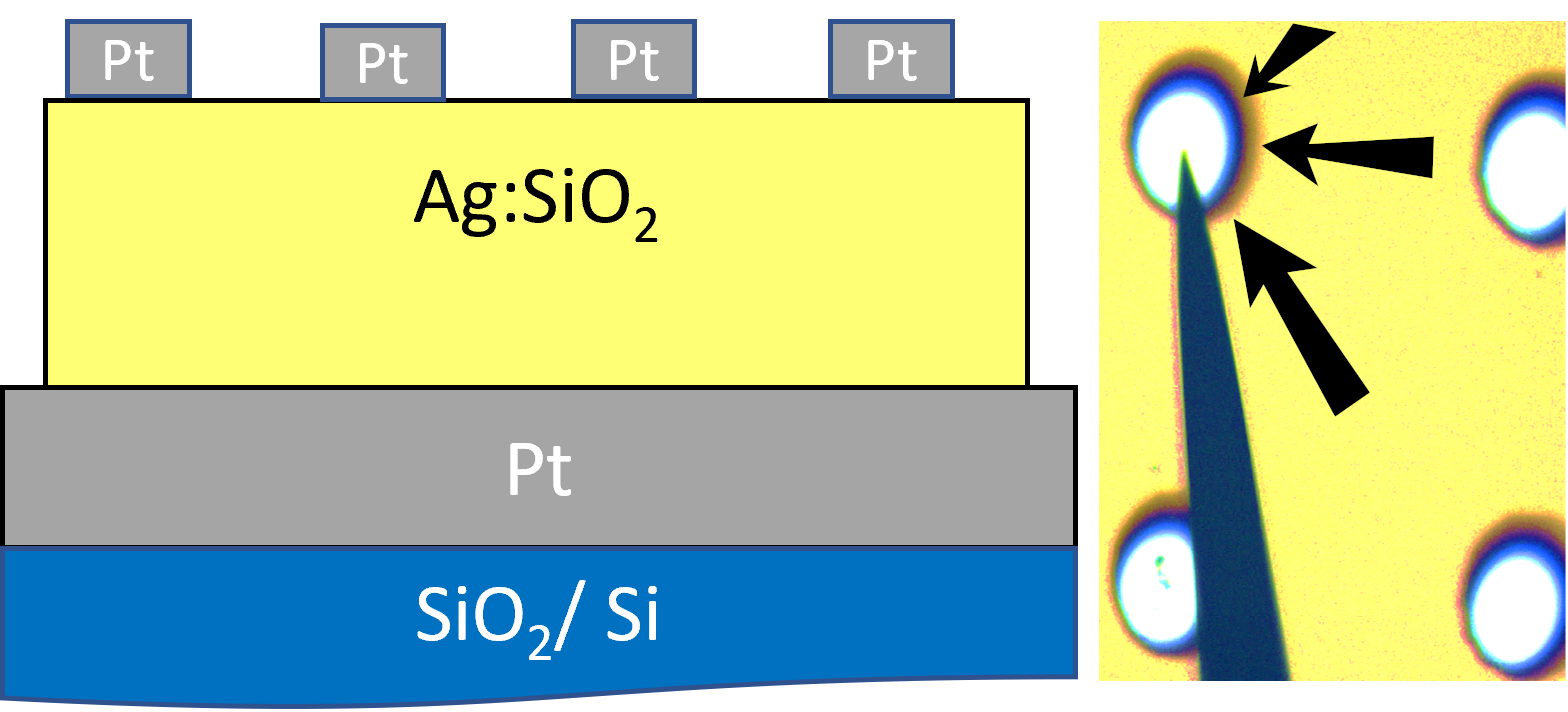}
	}
	\subfloat[] {
		\includegraphics[width=0.45\textwidth]{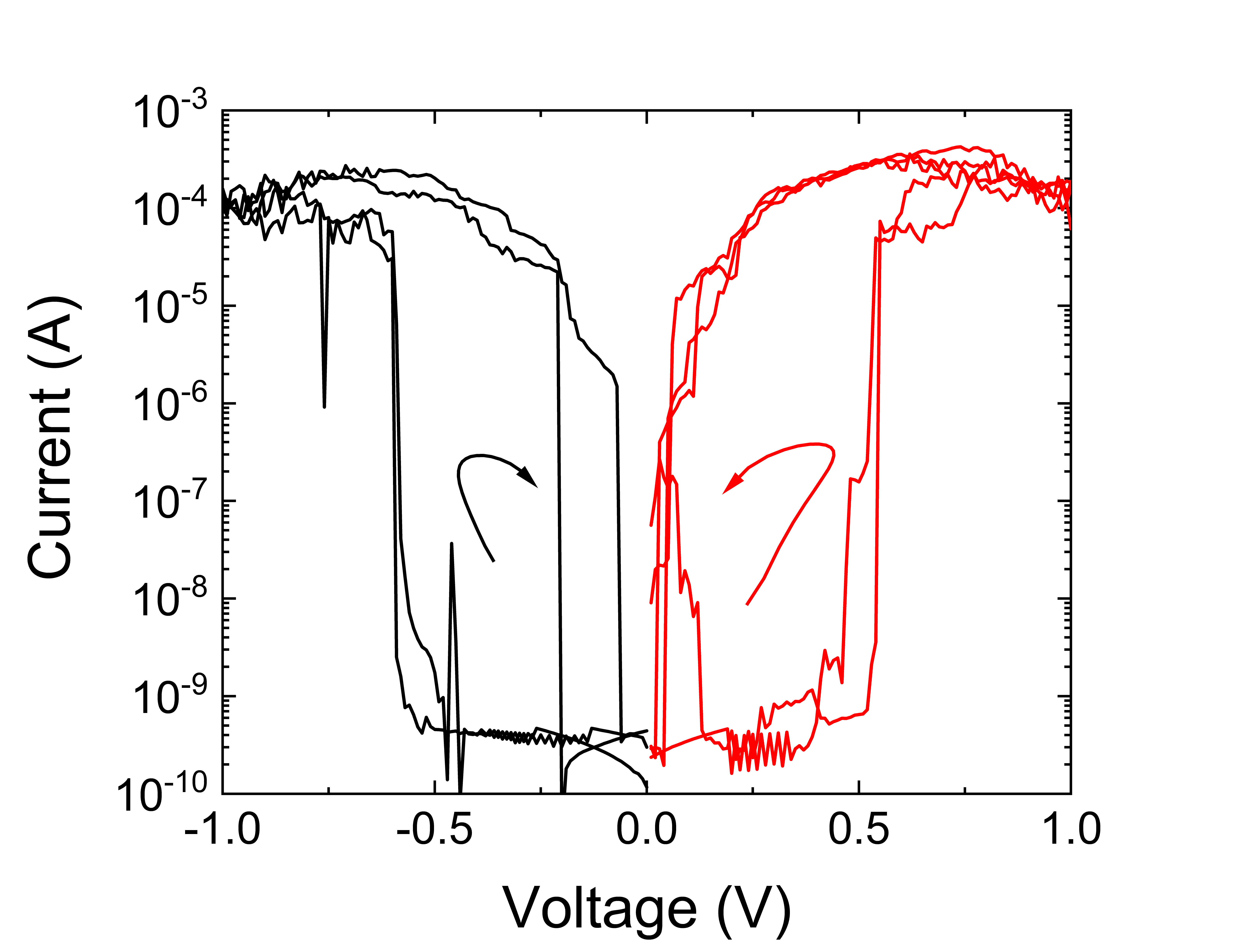}
	}
\caption{\label{fig:figure1}  (a) A schematic of a diffusive memristor device, Si/SiO$_2$/Pt(30 nm)/Ag:SiO$_x$(100 nm)/Pt (20 nm),
	and a top view photograph of the top Pt electrodes (100~$\mu$m diameter) with a probe tip placed for I-V measurements.
	Black arrows show possible diffusion directions of NPs to the space under the top electrode (see Sec.~\ref{sec:level2}).
	(b)Typical I-V characteristics of a device measured in both positive (red line) and negative	(black line) voltage cycles.
	The arrows indicate the direction of the change in device resistance. 
}	

\end{figure*}

\section{Sample details}

A schematic of the device structure is shown  in Fig.~\ref{fig:figure1}(a).
A platinum bottom electrode (30~nm) was deposited on a SiO$_2$/ Si(100)
wafer by magnetron
sputtering in a thin film deposition system from PVD Products Inc, with a fixed substrate to target distance of 15 cm. This was followed by co-sputtering Ag and SiO$_x$ using 7.6 cm diameter and 99.99$\%$ pure Ag and SiO$_2$ targets. The deposition was performed
in Ar and O$_2$ atmosphere at room temperature. The growth pressure was maintained
at 5~mTorr, and Ag and SiO$_2$ magnetron powers were set to 20~W (0.1~$\textit \AA$/sec) and 300~W (0.3~$\textit \AA$/sec) respectively.
To optimize the Ag concentration and obtain a relatively lower Ag deposition rate on our sample, the Ag target was masked using rectangular stainless steel racetrack that covered approximately 85$\%$ of the target area.
The top electrodes of Pt of thickness 20~nm (top part of Fig.~\ref{fig:figure1}a) were then sputter-deposited through a shadow mask with circular apertures of 100~$\mu$m diameter.

 Fig.~\ref{fig:figure1}(b) presents a typical dependence of the absolute values of the current through the memristor on the applied voltage for positive (red) and negative (black) voltage cycles. Both graphs evidence clear hysteretic behavior once the voltage magnitude first increases and then decreases, thus demonstrating  memristive properties of the device.  The I-V measurements were taken using a Keithley 4200 SCS with an attached Everbeing probe stage.

X-ray photoelectron spectroscopy (XPS)
	of the Ag~3d$_\frac{5}{2}$ and Si~2p was used to verify the silver concentration. XPS was performed using a Thermo K-Alpha system with an Al K$\alpha$ mono-chromated (1486.6~eV) source with an overall energy resolution of~350~meV. The analysis area captured was approximately 100~µm~x~200~µm. This was carried out in each corner of the sample and in the middle to check for homogeneity. For each location a survey was collected to preview the elements on the immediate surface, subsequent high-resolution scans were then performed on the elements of interest before fitting their peaks to identify elemental state.  All scans were charge corrected to adventitious C1s (C-C, C-H) peak at 284.8~eV. 	
The recorded XPS curves for the samples at different sites are shown in Supplementary figure S1 and table T1. \newline{}
 Transmission electron microscopes (TEM) specimens were prepared using the FEI Nova 600 nanoLab focused ion beam (FIB)/scanning electron microscope. A carbon layer was deposited on the film surface followed by a standard Pt deposition in the FIB as surface protection layers to preserve the outermost structure of the film. In addition, the carbon layer provides a gap between the FIB deposited Pt and the sputtered Pt top electrode to ensure accurate measurement of layer thickness and chemical analysis.

The TEM lamella had a thickness of approximately~100±20~nm for energy dispersive X-ray spectroscopy(EDS) and the tapered edge of~30~nm for high resolution imaging. TEM imaging and chemical analysis were performed using a FEI Tecnai F20 field emission gun scanning transmission electron microscope (FEGSTEM) equipped with a high angle annular dark field detector (HAADF) and Oxford Instruments X-Max 80 windowless X-ray detector for  EDS.

\begin{figure*}[htb!]
		\centering
		\subfloat[] {
			\includegraphics[width=0.42\textwidth]{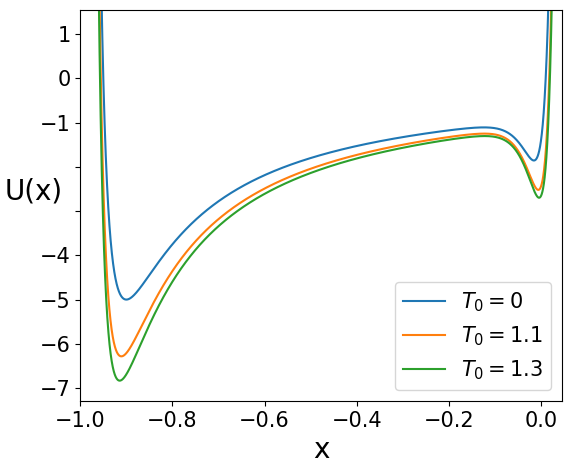}
		}
		\subfloat[] {
			\includegraphics[width=0.5\textwidth]{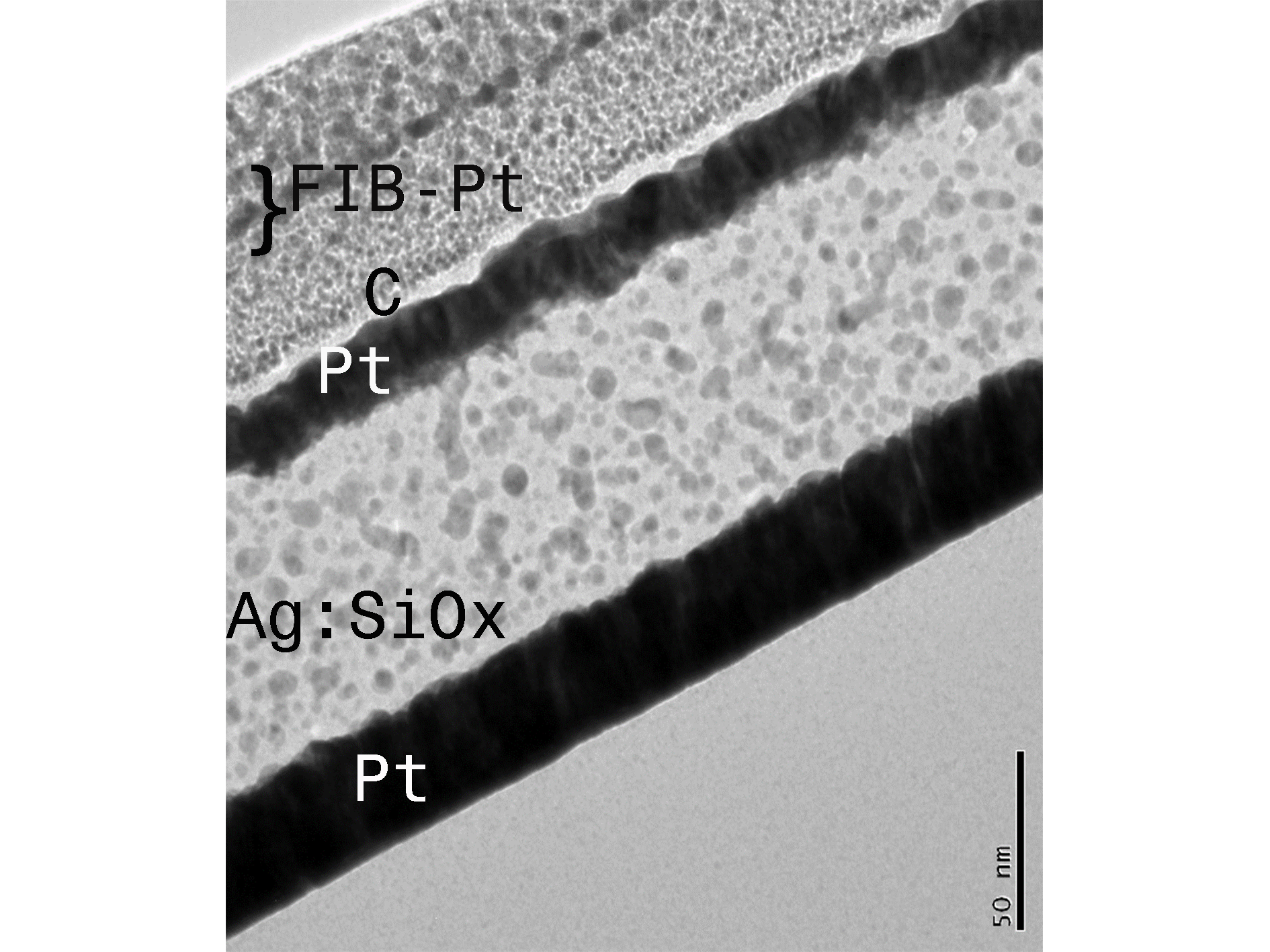}
		}
	\caption{\label{fig:figure2} (a) The normalized potential energy profile of a single Ag nanoparticle traveling between the two electrodes ($x=-1$ and $x=0$ correspond	to HRS and LRS of the memristor, respectively) as discussed in the model. The potential is modified with the increasing temperature. The normalization was done such that the local minimum near $x = 0$ is negative unit at $T_0 = 0$. (b) Cross-sectional TEM image of an as-grown sample. Darker Ag nanoparticles can be seen uniformly	distributed in the SiO$_x$ matrix. Thin white layer is carbon, the outermost two layers are FIB deposited Pt.}
\end{figure*}

\section{\label{sec:level3}Theoretical model}

To qualitatively describe the resistive switching in a diffusive memristor
we utilized a modified phenomenological model that previously explained the experimental findings in
Refs.~\cite{savel2013mesoscopic,wang2017memristors,jiang2017novel,ushakov2021role, ushakov2021deterministic}.
The main dynamical equation is
\begin{equation}\label{eq:model_1}
	\dot{x} = -U'(x) + V/2 + \sqrt{T}\xi(t),
\end{equation}
where an effective potential $U(x)$ accounts for the interaction of NP with various inhomogeneities and other NPs as well as various electrochemical potentials and temperature gradients. We also introduce notations $q, V, T,$ and $\xi(t)$ representing the particle charge, voltage applied to the memristor terminals, nanoparticle temperature, and zero-mean $\delta$-correlated white Gaussian noise ($\langle \xi\rangle=0$, $\langle\xi(0)\xi(t)\rangle=\delta(t)$), respectively. Time and coordinate derivatives are denoted by $\dot{x} \equiv dx/dt$ and $U' \equiv \partial U/\partial x$. 
The position of the nanoparticle defines the resistance of the device, which is approximated as $R(x) = \cosh(x/\lambda)$ with all resistances normalized by the lowest resistance $R_0$ of the memristor. This approximation implies exponential electron tunneling through the particle between memristor terminals, where $\lambda$ is the tunneling length~\cite{jiang2017novel}. Instead of modeling many diffusing NPs and estimating the device resistance by considering many competing paths connecting the top and the bottom memristor electrodes through NPs, we assume that the resistance of the modeled memristor is actually determined by a bottleneck of size $L$ linking two low-resistance clusters, such as two parts of almost formed CF (see, e.g., Ref.~\cite{yi2016quantized}). In this case, $x$ can be interpreted as a position of the single NP determining resistance of the bottleneck. Under these circumstances, the shape of $U(x)$ can be hardly evaluated from the microscopic calculation and may be influenced by the distribution of particles around the bottleneck, which in turn depends on many factors including bath temperature $T_0$. Here we have examined several potentials and concluded that the main features of the memristor's switching are robust against details of the potential which can be chosen phenomenologically after the analysis of the TEM images of the device, and I-V measurements. Note that in the dimensionless equation~(\ref{eq:model_1}), the time is normalized by $t_{norm}=\eta L^2/\Delta U$ with the viscosity $\eta$ of Ag clusters in SiO$_2$ matrix and the depth $\Delta U$ of the potential well corresponding to LRS of the memristor (see the caption of Fig.~\ref{fig:figure2}), the voltage is normalized by $V_{norm}=2 \Delta U/q_0$ with the Ag cluster charge $q_0$, and the temperature is normalized by $T_{norm}=\Delta U/2$. For the dimensionless temperature, note that it corresponds to the dimensionless energy (discussed in Supplementary materials).

In contrast to the previous works,
here we used a double-well potential (Fig.~\ref{fig:figure2}(a)) to reflect the observed stability of the device both in HRS and LRS. Within the framework of our model,  the position of the nanoparticle in the vicinity of  $x = -1$
corresponds to HRS of the device, while the one in the vicinity of $x = 0$ defines LRS. This suggests that the two wells of the potential $U(x)$ should reside at $x=0$, and $x=-1$ standing for HRS (note that we also can select $x=1$ for HRS if simultaneously change $V$ to $-V$).

The nanoparticle temperature $T$ in Eq.~(\ref{eq:model_1}) is a dynamical variable depending on $V$ and
	$R(x)$ and obeying the equation
\begin{equation}\label{eq:model_2}
	\dot{T} = \frac{V^2}{C_h R(x)} + k(T - T_0),
\end{equation}
where $C_h$ and $k$  denote the heat capacity of the device and heat transfer coefficient, respectively.
 The equation~(\ref{eq:model_2})
	reflects the famous Newton cooling law describing heat dynamics due to the Joule heating and heat transfer to the substrate.
As was shown in  Refs.~\cite{ushakov2021role,ushakov2021deterministic}, the interplay of heat, Ag-cluster diffusion and electron transport can result in a very rich dynamical features in the diffusive memristor.

We selected the dimensionless parameter values as well as the shape and temperature dependence of the potential so that to provide a qualitative agreement between the theory and experiment (see the exact expression in Supplementary materials).

\begin{figure*}[!htb]
	\centering
	\subfloat[] {
		\includegraphics[width=0.4\textwidth]{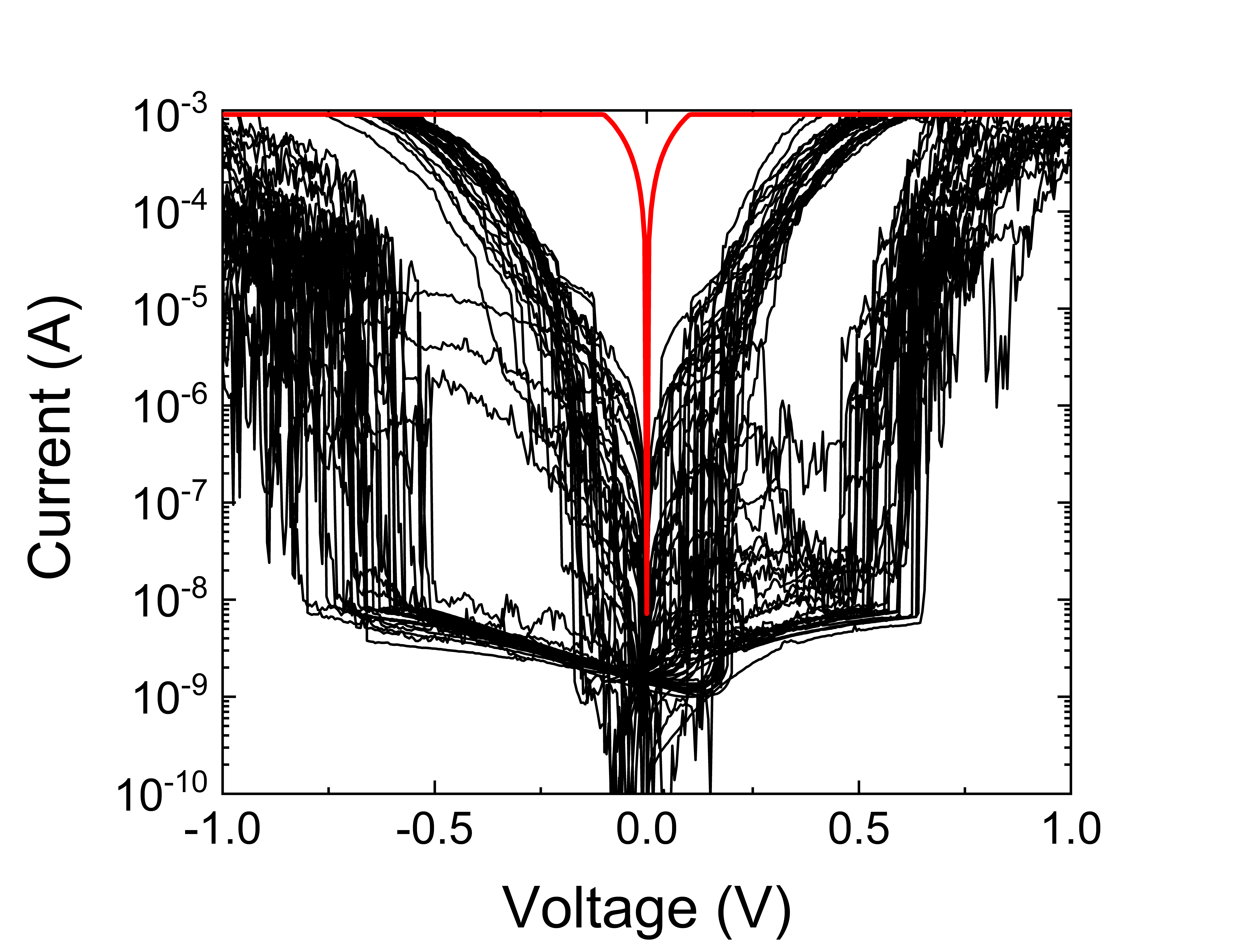}
	}
	\subfloat[] {
		\includegraphics[width=0.33\textwidth]{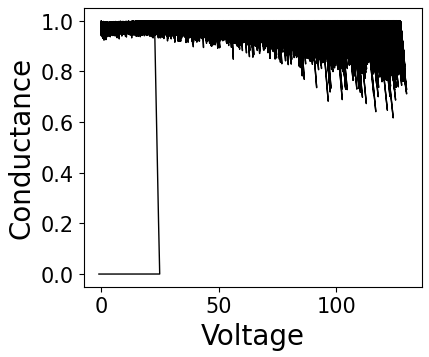}
	}

	\subfloat[] {
		\includegraphics[width=0.3\textwidth]{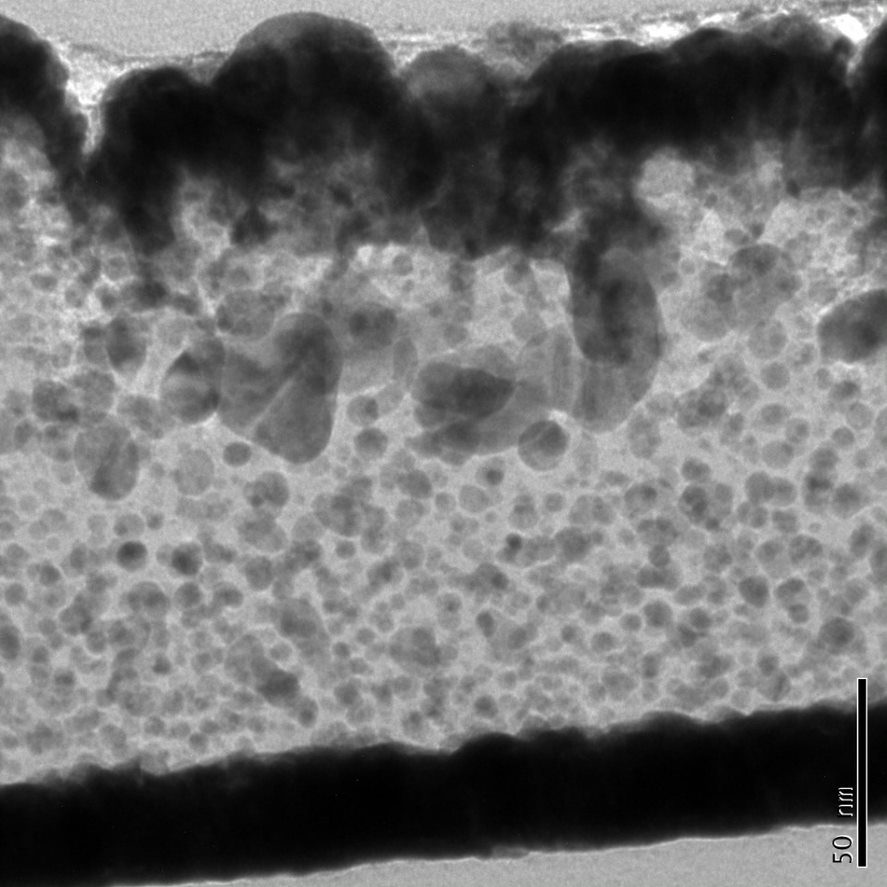}
	}
	\subfloat[] {
		\includegraphics[width=0.3\textwidth]{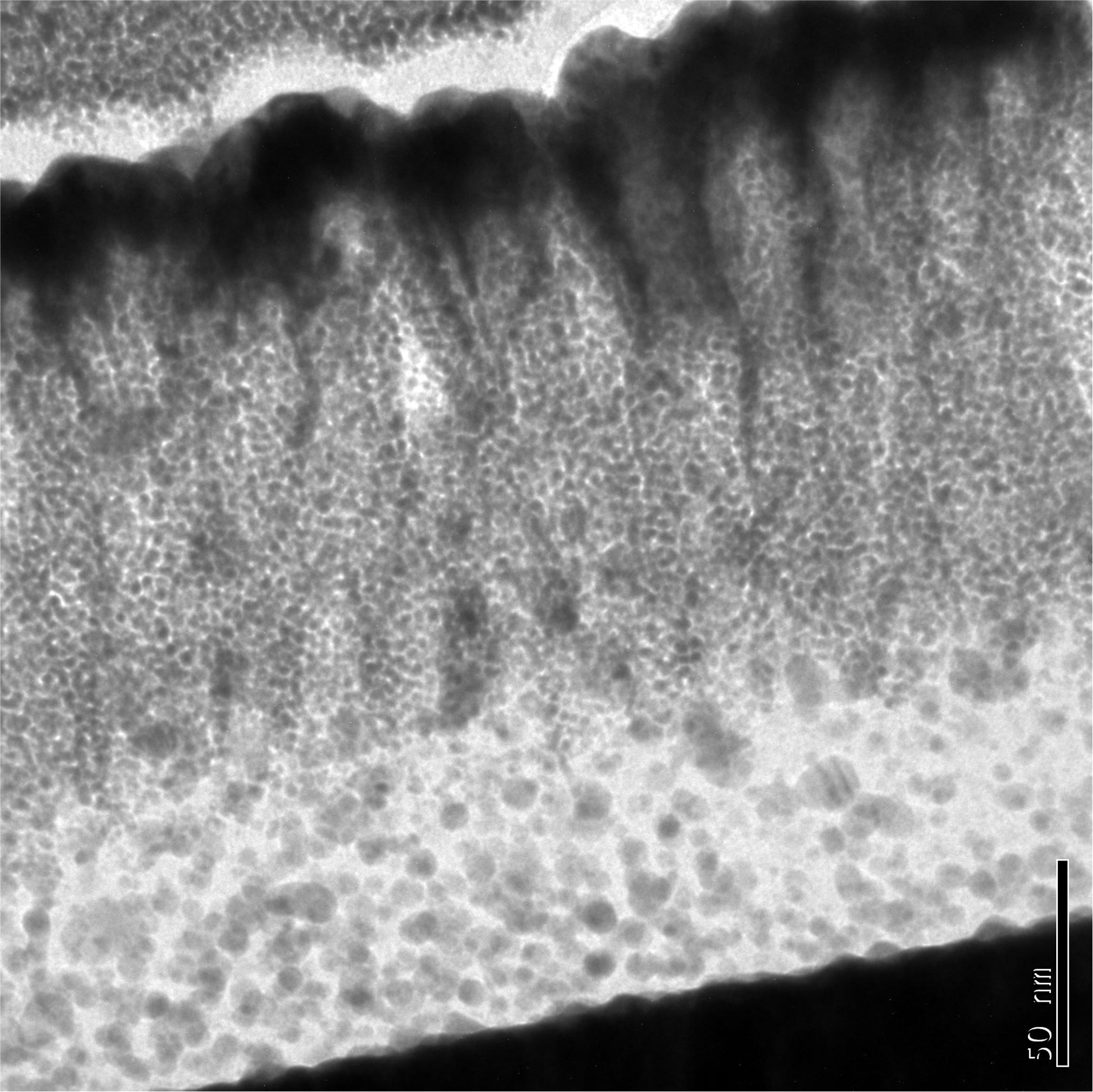}
	}
	\caption{\label{fig:figure3} (a) Experimental I-V curve. Initial I-V switching (black lines) transforms into a permanent
		LRS (red line) after multiple voltage cycles. (b) Simulated behavior of the conductance ($1/R(x)$) of a diffusive memristor
        in response to external voltage.
		When the NP is trapped in the metallic state at position $x = 0$
		(see Fig.~\ref{fig:figure2}a), the device
		conductance stays high ($T_0 = 0.1$). 
		(c) Cross-sectional TEM images of samples at room temperature when in permanent LRS, and (d) after the substrate
		temperature was raised to 50$^o$C to break the Ag clusters at the top electrode site.
	}
\end{figure*}

\section{Results and discussion}

\subsection{\label{sec:level2} Memristor reset.}

 Fig.~\ref{fig:figure2}(b) presents a bright field TEM image of the cross section of an as-grown device before applying any voltage cycles to the Pt electrodes. Silver nanoparticles  (dark gray dots)
	can be observed rather uniformly distributed in the SiO$_x$ dielectric matrix sandwiched between the top and bottom Pt electrodes. A typical I-V hysteresis during application of an external voltage to the top and bottom Pt electrodes at the room temperature (Fig.~\ref{fig:figure1}(b)) demonstrated a volatile switching behavior for both voltage polarities.
 
	After  a  number of voltage cycles where the device showed distinct I-V hysteresis switching
	(Fig.~\ref{fig:figure3}(a), black lines), the memristor switched into permanent LRS.
	Subsequent voltage cycles, either positive or negative, did not reset the device to HRS (see red curves in  Figs.~\ref{fig:figure3}(a), which do not demonstrate hysteresis anymore). Analysis of TEM images (Figs.~\ref{fig:figure3}(c)) revealed that this state of the device is typically characterized by the presence of relatively large silver clusters near	the top electrode and densely packed Ag NP near the bottom one. 
	
	To understand such behavior,  let us  consider qualitatively the mechanisms behind the filament formation in the device. When an external voltage is applied to the memristor terminals, the generated Ag$^+$ ions start to drift towards the bottom negative electrode~\cite{yang2014electrochemical,jimenez2007optical}.
	As such field-dependent drift takes place under the top Pt electrode (shown by arrows in Fig.\ref{fig:figure1}a), the neighbouring neutral Ag nanoparticles diffuse to the region of
	the lower Ag concentration. They come from the space around the top electrode. Indeed, one can see in Fig.~\ref{fig:figure1}(a) that the top electrode is a small spot over the Ag:SiO$_2$ layer, so the strongest directed
	electric field takes place under that small spot only.
	Fundamentally, this process leads to a CF formation. When the value of the applied voltage decreases,
	the CF breaks due to thermal fluctuations and diffusion processes, so the memristor switches back to HRS. 
	Similar mechanisms are realized also for opposite voltage polarity.
	
	Multiple repeating cycles of formation and rupturing of the CFs, which are highly stochastic processes,  can eventually  lead to development of an asymmetry in the particle distribution.
	In our case, it manifests itself in asymmetric clustering near the top and bottom electrodes. Namely, while the vicinity of the top electrode is occupied by rather large clusters of Ag-nanoparticles (spatial dark volumes), the bottom electrode accumulates much lesser fraction of the NP clusters (Fig.\ref{fig:figure3}c).
	Such  asymmetry could be explained  by the difference in concentration of the pinning sites on 
	the top and bottom electrodes, which is caused by the higher roughness of the top
	electrode~\cite{yang2014electrochemical, zhao2020strategy,onofrio2015atomic}, as evident from the TEM images
	(see Fig.~\ref{fig:figure2}(b) and Figs.~\ref{fig:figure3}(c,~d)).  Appearance of such pinning centers, which attract and trap nanoparticles, is conditioned by a number of factors including  the surface tension, cohesion forces, and the Rayleigh instability~\cite{wang2018threshold}. The  NP clusters at the top serve
	as an extension of the top electrode making the actual width of the memristor narrower, thus, the short-circuiting CF formation may occur
	with higher probability.

\begin{figure}[h]
	\subfloat[] {
		\includegraphics[width=0.4\textwidth]{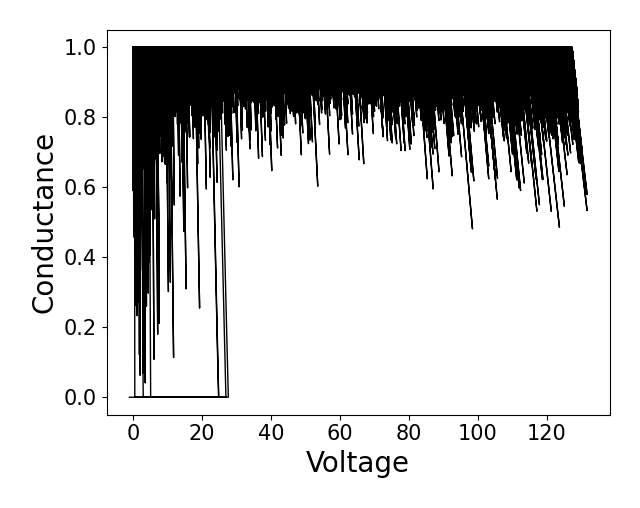}
	}

	\subfloat[] {
		\includegraphics[width=0.4\textwidth]{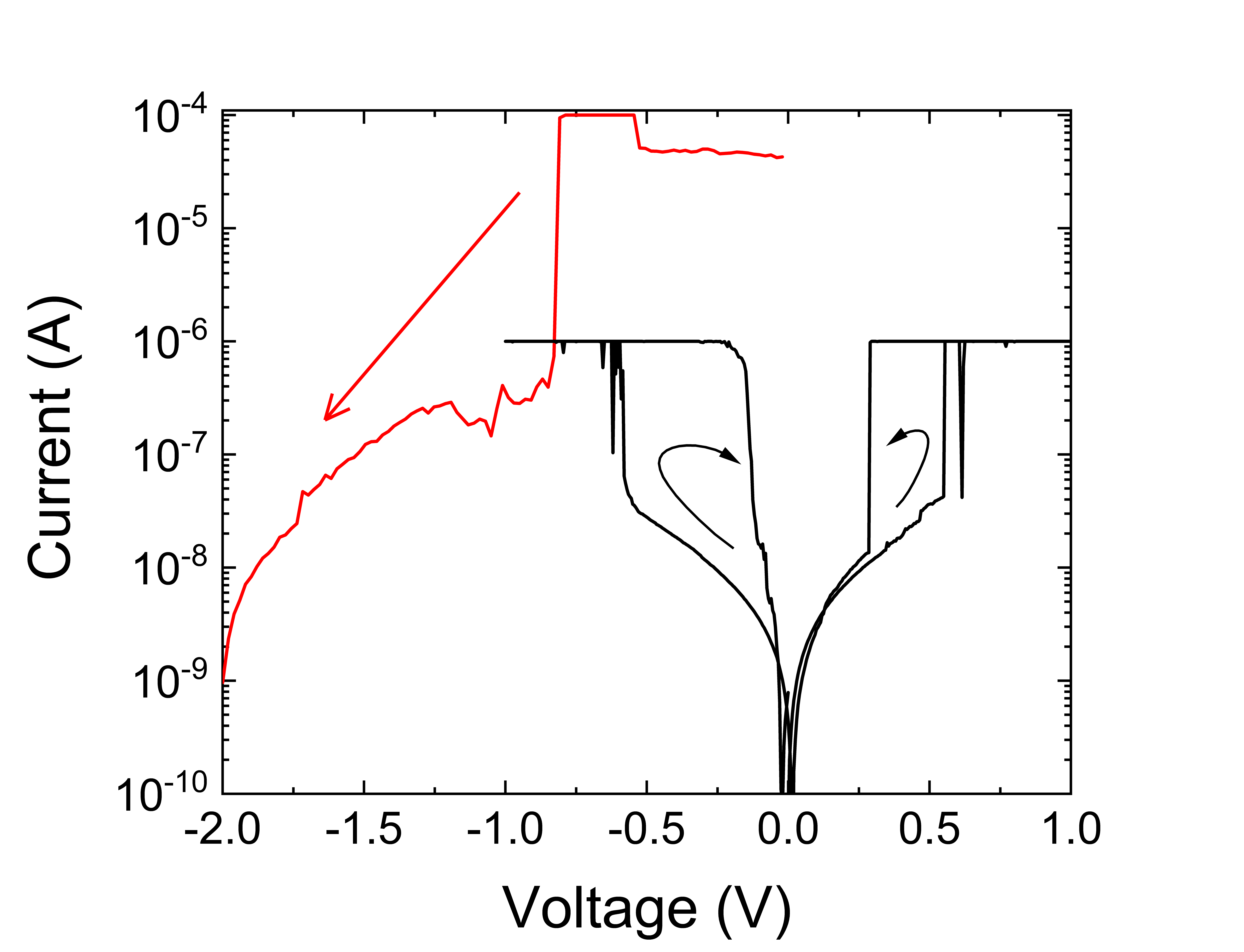}
	}
	\caption{\label{fig:figure4} (a) Simulated memristor conductance  and noise optimisation behavior at raised device
		temperature ($T_0 = 1.5$). A volatile switching between LRS and HRS is observed at $V \approx 30$ and at $V \approx 0$
		(cf. Fig.~\ref{fig:figure2}a).
		(b) Experimental I-V curves demonstrating the first electroforming sweep assisted by external heat (red line) by hosting the sample on a temperature controlled stage at 50$^o$C. 
	}
\end{figure}

	To get a deeper insight in the mechanisms setting the memristor to the permanent LRS we employed the model  (\ref{eq:model_1}), (\ref{eq:model_2}). We considered both temperature independent $U(x)$ and temperature dependent $U(x, T_0)$ (see Fig. \ref{fig:figure2}(a)). From our simulations, we demonstrated that the reset to HRS Figs.~\ref{fig:figure3},~\ref{fig:figure4} can be described by using the simple two-well $U(x)$, however, the temperature dependence of switching and transformation of spiking with temperature (see subsections~\ref{sec:temp_depend},~\ref{sec:spiking}) can be described by assuming temperature dependence of the potential $U(x, T_0)$. Asymmetry in clustering demonstrates the asymmetry of potential wells with the deepest well near  $x=-1$, which is closer to the top electrode. Under this consideration, the LRS state should correspond to the situation when NP arrives to the vicinity of the potential minimum near $x=0$, where $R(x)$ reaches its lowest value.
	
	For the given shape of the potential, at $T_0$=0.1, the model reproduced the experimental situation as illustrated by the conductance (inverse resistance) curve in Fig.~\ref{fig:figure3}(b). The system starts from HRS and then eventually switches to LRS as the applied voltage increases. Then, any further increases and decreases of the voltage do not produce any hysteresis in the conductance curve, and the corresponding curves although becoming noisy demonstrate rather the same path. Such a behavior of the model suggests that NP is trapped in the potential well corresponding to LRS. We hypothesized that additional energy stimulating particle escape from the potential minimum could be pumped in by increasing the temperature of the device $T_0$. At least, in the corresponding TEM image (Fig.~\ref{fig:figure3}(d)), which was taken for the sample after temperature annealing, one can observe that the mentioned top silver clusters have been broken into small ones. In other words, the increased temperature improved the agility of NPs.

The hypothesis can be easily verified in simulations. Fig.~\ref{fig:figure4}(a) presents resetting of the hysteretic behavior as  $T_0$ was set at $1.5$.
At this temperature, changing of voltage returned the system to HRS and restored a hysteresis in the I-V curves.
	
To test the hypothesis in practice, a sample in the permanent LRS was hosted on a temperature-controlled
sample stage, and the stage temperature was gradually increased to 50$^o$C. A negative voltage sweep was first
applied followed by a complete positive and negative voltage cycle. The negative voltage sweep (electroforming)
assisted by the external thermal contribution  pushed the filaments from LRS  to
HRS. When the voltage was cycled at low temperature again, the device demonstrated I-V hysteresis
characteristic (Fig.~\ref{fig:figure4}(b)). It is seen that during electroforming to negative voltages at 50$^o$C, the device current dropped to low values (shown by red line), and when a subsequent full voltage sweep was performed, the memristor started from HRS and showed a hysteresis current voltage characteristic (shown by black lines, with the arrow representing the direction of resistance switching).

Besides broken top silver clusters, the net Ag$\%$ at the surface was also shown to increase (see supplementary table T1), and there was a shift in the Ag 3d peaks. The shift in the metallic Ag peaks can be correlated with the size difference of NPs~\cite{chiappone2018situ}.

 It is worth to note, a further theoretical analysis showed that decreasing the thermal conductivity or the specific heat capacity of the memristor could also reset the permanent LRS. For example, the conductance behavior has been modeled for different values of the specific heat capacity $C_h$ and the heat transfer coefficient $k$ using  Eq.~(\ref{eq:model_2}), and shown in Supplementary figure S2.  However, implementation of such control requires complex design of the device which is beyond the scope of this paper.
 
To understand the reason why increasing $T_0$ or decreasing $C_h$ or $k$ allows for resetting the HRS, one can analyze switching conditions. To switch to LRS the system should escape from the potential well $x=-1$ corresponding to HRS. When voltage increases (assuming $V$ being positive), as soon as the force $V/2$ exceeds the maximum restoring force  $V_c/2 = \underset{-1<x<0}{\max} |\partial U/\partial x|$, the NP diffuses to $x=0$ (LRS). When voltage decreases, the particle remains trapped at the well $x=0$, and only temperature fluctuations can push the particle to hop over the barrier back to HRS. Therefore, the bath temperature $T_0$ should be about the barrier height, which is $T_0 \approx 1$ in the chosen dimensionless units.

\subsection{\label{sec:temp_depend}Dependence of switching threshold upon temperature}

\begin{figure*}[ht]
 	\centering
 	\subfloat[] {
 		\includegraphics[width=0.45\textwidth]{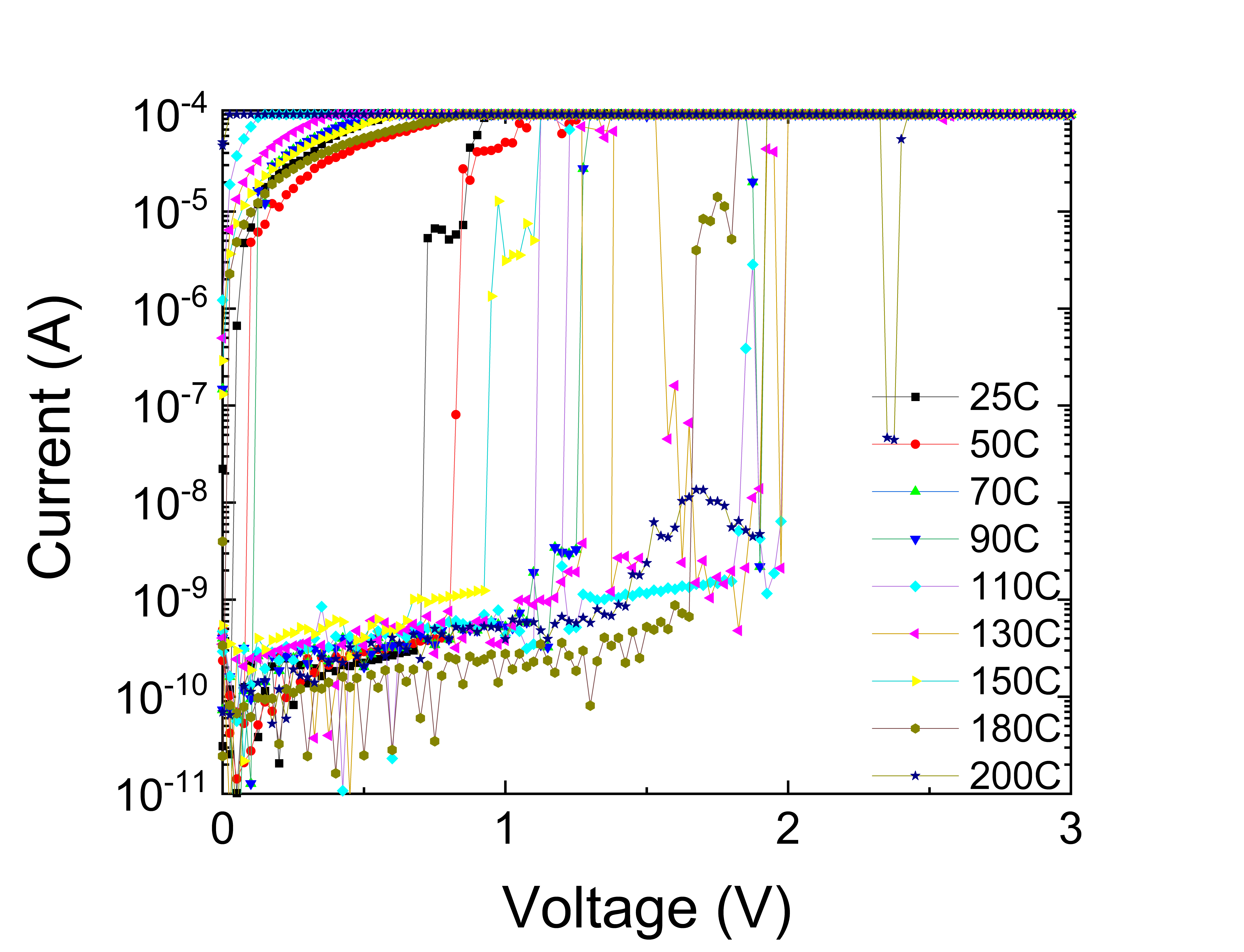}
 	}
 	\subfloat[] {
 		\includegraphics[width=0.4\textwidth]{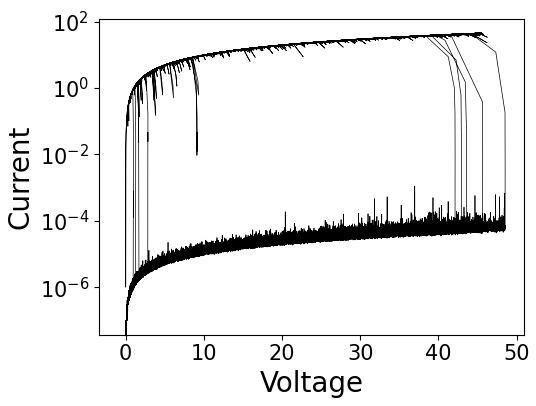}
 	}
 	
 	\subfloat[] {
 		\includegraphics[width=0.45\textwidth]{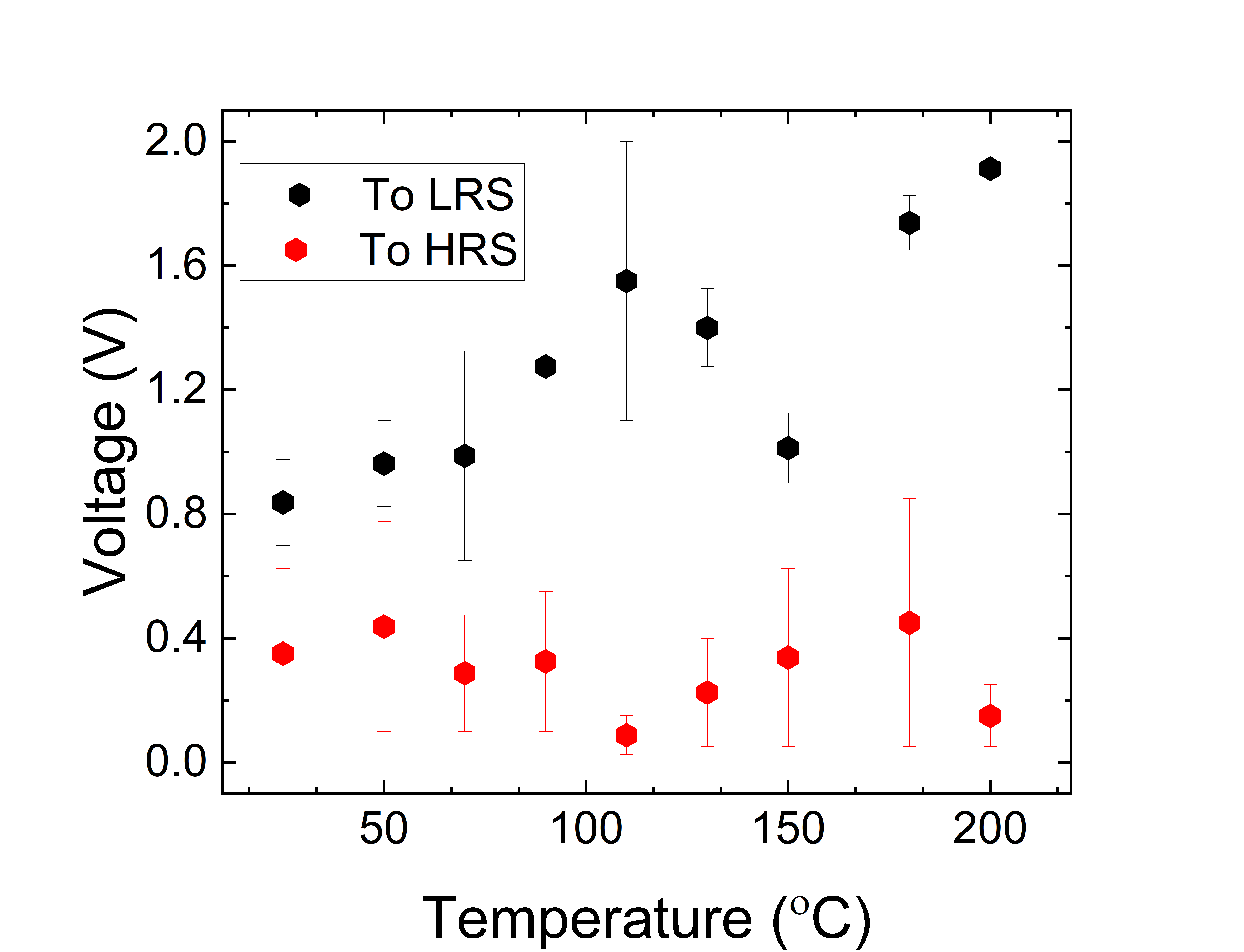}
 	}
 	\subfloat[] {
 		\includegraphics[width=0.38\textwidth]{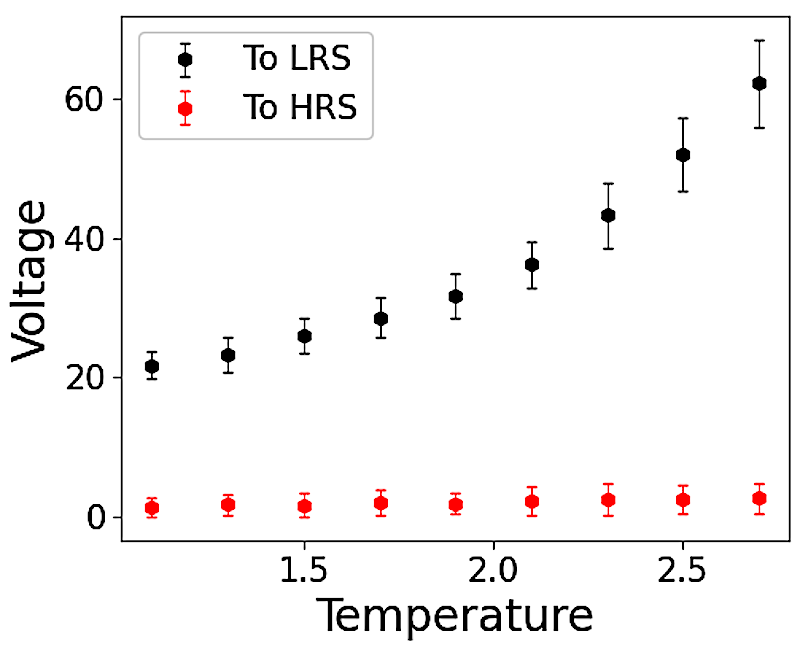}
 	}
 	\caption{\label{fig:figure5} (a) Experimental I-V loops measured at different hold temperatures showing distinct threshold voltage for HRS to LRS transition.
 		(b) Simulated I-V loop at $T_0 = 2.3$ for temperature dependent potential shown in Fig.~\ref{fig:figure2}. (c) Experimental and (d) Simulated variation of switching voltage for HRS to LRS (Black) and LRS to HRS (Red) switch for the potential and parameters as in (b).
 	}
 \end{figure*}

 Our experiments demonstrated the possibility of the memristor reset in a range of temperatures. Fig.~\ref{fig:figure5}(a) summarizes the measured I-V curves  after reset for $T_0$ between 25$^o$C and 200$^o$C. Remarkably, as $T_0$ varies the voltage thresholds for switching the memristor  to HRS and LRS change.  Fig.~\ref{fig:figure5}(c) presents how the averaged over 5 cycles threshold voltage and  the respective error bars for switching to LRS (black) and to HRS (red) change with variation of the device temperature.  While the averaged voltage to switch to LRS significantly grows with increase of $T_0$, the threshold voltage for setting the memristor to the HRS weakly changes. Such trends in the thresholds dependencies on temperature of the device could be explained by the reasonable assumption that the higher temperature intensifies the random fluctuations of NPs. Thus, the system requires higher voltages to form a conducting filament out of smaller particles,
 not completely shrunk top NP clusters, distance between the terminals (see Fig.~\ref{fig:figure3}(d)).
 
 Within the framework of our model, this suggests a temperature dependent potential $U(x, T_0)$
 with deeper wells for higher $T_0$, see Fig.~\ref{fig:figure2}(a). The fact of a weak temperature dependence of the thresholds for switching to HRS could be explained  by sintering~\cite{magdassi2010triggering} of NPs, which takes effect at higher temperatures. This implies that as soon as CF gets formed it remains stable despite the increased thermal fluctuations,
 so the model particle gets trapped in a deeper potential minimum $x = 0$ (lowest R(x) value). 
 Our theoretical model with the potential depending on $T_0$ demonstrates a good qualitative agreement with the results of experimental measurements. In particular, it confirms possibility of resetting the memristor from a permanent LRS to HRS within a range of the temperature. For example, Fig.~\ref{fig:figure5}(b) presents hysteretic behavior of I-V curves for $T_0=2.3$. The calculated dependencies of the  voltage thresholds for switching to LRS and HRS (Fig.~\ref{fig:figure5}(d)) reproduce well the trends of experimentally measured dependencies (cf. Fig.~\ref{fig:figure4}(b)).
 
\subsection{\label{sec:spiking}Temperature effect on a spiking circuit}
\begin{figure*}[htb]
    \subfloat[] {
        \includegraphics[width=0.3\textwidth]{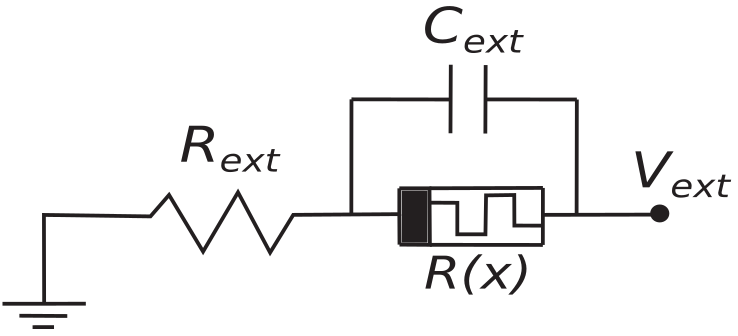}
    }

	\subfloat[] {
		\includegraphics[width=0.4\textwidth]{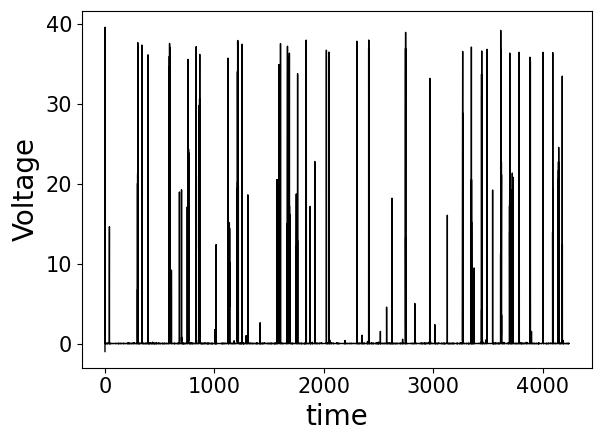}
	}
	\subfloat[] {
		\includegraphics[width=0.4\textwidth]{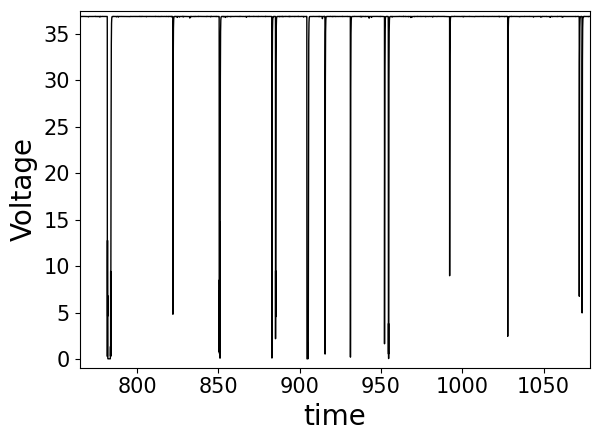}
	}
	
	\subfloat[] {
		\includegraphics[width=0.4\textwidth]{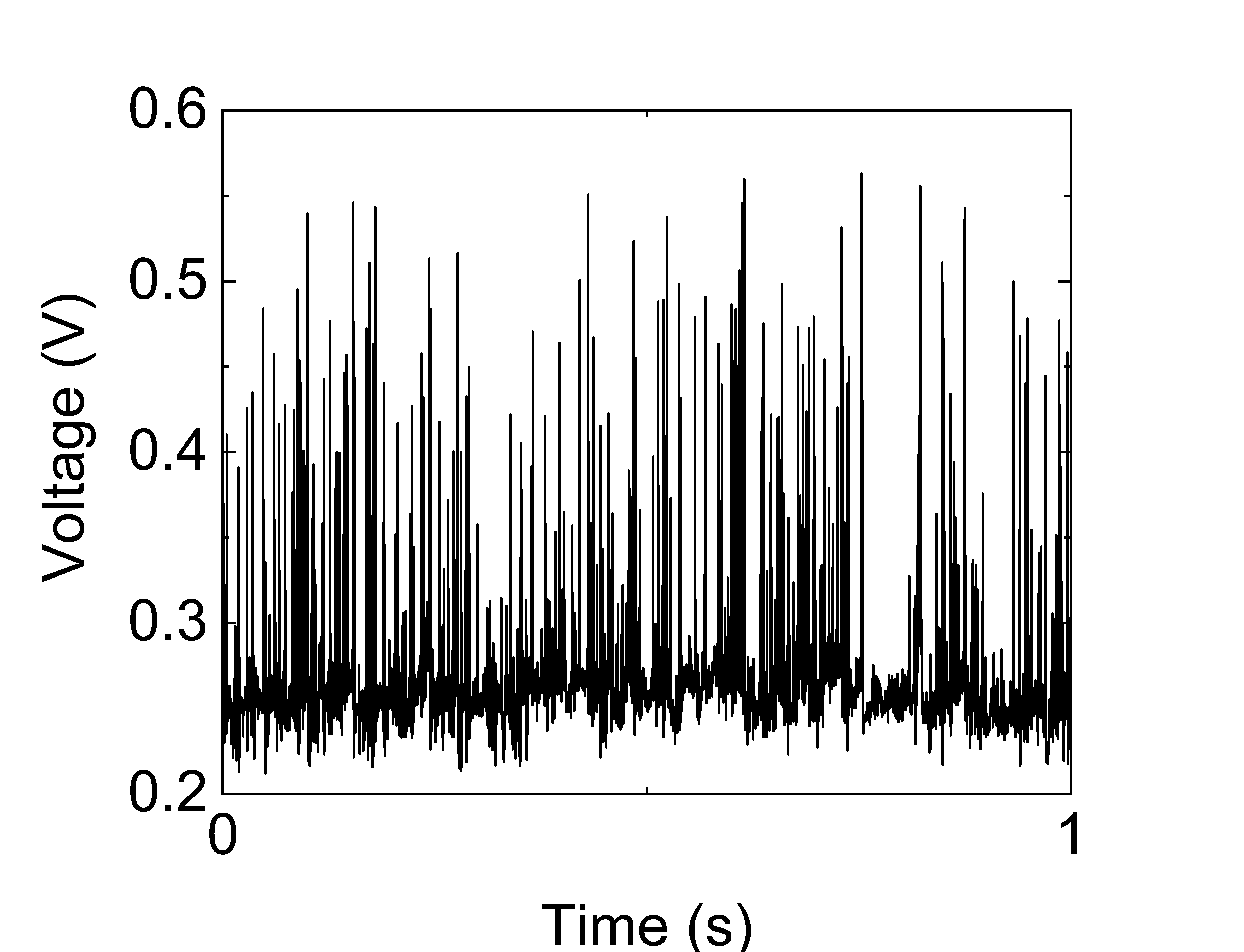}
	}
	\subfloat[] {
		\includegraphics[width=0.4\textwidth]{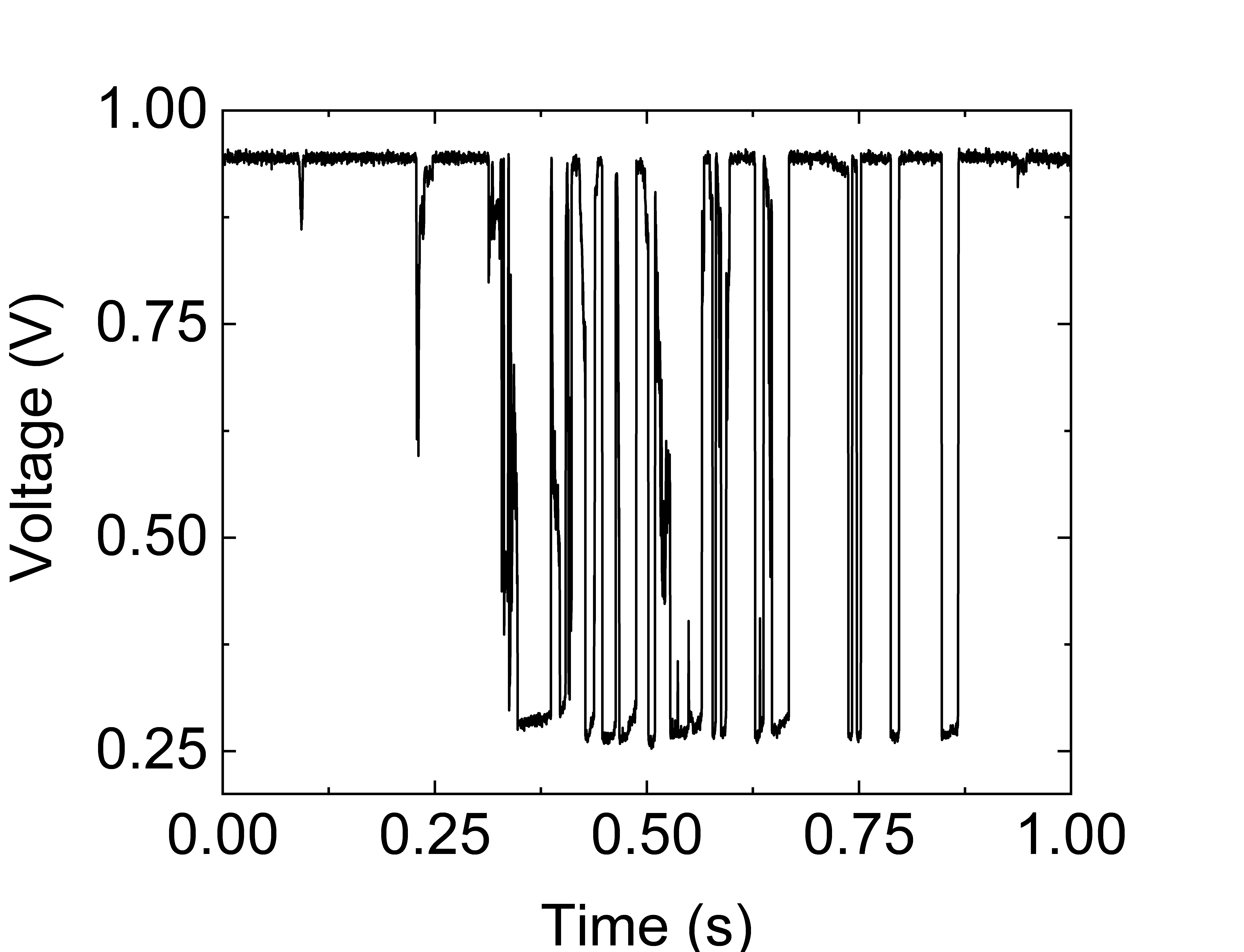}
	}
	
	\caption{\label{fig:spiking}
        (a) Artificial neuron circuit.
		(b) Simulated bottom-up voltage spikes of (a) at low temperature ($T_0$ = 1.1) for the parameters and potential as in Fig.~\ref{fig:figure5}.
		The voltage is mainly at the low level, which corresponds to LRS of the memristor.
		(c) Simulated top-down voltage spikes at higher temperature ($T_0 = 2.1$s), that is, the memristor is mostly in HRS, for the parameters and potential as in Fig.~\ref{fig:figure5}.
		(d) Experimental voltage spikes for the diffusive memristor, when SiO$_x$ matrix is set to room temperature 
		and (e) the temperature of 200$^o$C.
	}
\end{figure*}

 One of the most promising application of diffusive memristors is using them as a core element in spiking and switching neuromorphic devices ~\cite{jeong2012emerging,yang2013memristive,sangwan2020neuromorphic}. Therefore, next we investigated how the change of memristor temperature $T_0$ affects the spiking regimes of an artificial neuron, whose principal electrical scheme is depicted in Fig.~\ref{fig:spiking}(a). Here, $R_{ext}$ is a load resistor and $C_{ext}$ describes internal or/and externally added capacitance of the memristor.
 According to the Kirchhoff's law, in this case,  the voltage drop over the memristor, $V$, obeys the equation
 \begin{equation}\label{eq:V}
 	\tau \dot{V} = V_{ext} - \left(1 + \frac{R_{ext}}{R(x)}\right)V,
 \end{equation}
 where $V_{ext}$ is the externally applied constant voltage and $\tau = R_{ext}C_{ext}$.
 Numerical simulation of the Eqs.~(\ref{eq:model_1}-\ref{eq:V}) reveals that change of $T_0$ can regulate the residual time the system spends in HRS and LRS during spiking cycles. Figs.~\ref{fig:spiking}~(b,~c) present the time-realization of $V$ for $T_0$=1.1 and $T_0$=2.1, respectively, at the same $V_{ext}$=37. It is clearly seen that for $T_0$=1.1~(a), the state where the memristor spends more time corresponds to LRS (low voltage background), and the spike manifests itself as a sharp surge of $V$ towards its higher values followed by a swift return to LRS. An opposite picture is observed for larger temperature $T_0$=2.1~(b), where the $V$-spikes are associated with sharp drop of the voltage over the memristor followed  by the return to higher voltage values related to HRS.  We note that changing the state, where the system spends more time, from LRS to HRS with variation of $T_0$ relates to changing the stability properties of the equilibrium points of the dynamical system (\ref{eq:model_1}-\ref{eq:V}), whose analysis will be reported elsewhere.

 To check our numerical findings in experiment, we connected an external resistor $R_{ext}$ in series to the memristor discussed in Sec.~\ref{sec:level2}, assuming the presence of an internal memristor capacitance. The voltage across the device was measured using an oscilloscope at different voltage amplitudes. The value  of  $R_{ext}$ was fixed at 55~k$\Omega$, which produced spiking in a wide range of voltages. The dynamics of the voltage drop over the memristor for two different temperatures is illustrated by Figs.~\ref{fig:spiking}~(d) and (e). For the room temperature (d), and $V_{ext} = 0.6$~V corresponding to threshold voltage at RT, the circuit demonstrated spiking similar to one presented in (b), i.e. where spikes are associated with sharp increase of $V$. However, as the temperature was set at 200$^o$C, the spiking regime occurred for higher voltages.  Moreover, as  Fig.~\ref{fig:spiking}~(e) evidences for $V_{ext} = 1$~V, the spiking was inverted similarly to our numerical simulations (see Fig.~\ref{fig:spiking}(c)). Also, as in the numerical simulations, in this case the most time the system spends in HRS and only occasionally visits LRS.  The possibility to control time the system spends in HRS or LRS by setting the appropriate temperature offers another degree of freedom in controlling artificial neuron dynamics.

\subsection{\label{sec:level6}Device structural damage at higher temperatures}

 Having proposed temperature as a mean to control the functioning of the diffusive memristors, it is important to find a proper range of the temperature tuning. In our experiments we observed that a long-time exposure of diffusive memristors to high temperature can cause a permanent structural damage.
The relevant TEM/SEM images of a prolonged heat treated sample were taken and are shown in Supplementary figure S3.
It was revealed that with extended heating time the nanoparticles coalesce together and surface out of the sample. This property is very similar to Ag sintering properties and can be attributed to Ostwald ripening of Ag
nanoparticles~\cite{magdassi2010triggering, jimenez2007optical, yeo2014spin}.

  \section{Summary and Conclusions}

 In this work we experimentally and theoretically demonstrated that tuning of the temperature of the dielectric matrix of a diffusive memristor can be used as efficient tool to control hysteretic properties of the device and spiking regimes in the memristor based artificial neuron circuit. In particular, we showed that appropriate temperature tuning can restore the hysteresis in I-V characteristics of a diffusive memristor which gets stuck in LRS after a number of operating cycles. This provides a way to increasing the life time or endurance of the diffusive memristor devices. On the other hand, the temperature reset offers an approach to manipulate the hysteretic properties of the device by giving a possibility to switch between hysteretic and non-hysteretic behavior of the memristor resistance.  
	
	Another finding showed the possibility to employ the matrix temperature in order to regulate the voltage thresholds for switching between HRS and LRS. We found out that while the voltage threshold for switching to LRS strongly depends on the temperature, the threshold for switching to HRS is less sensitive to the temperature change. These results suggest using the temperature for tuning the hysteresis width, which is important both for resistive switches and for spiking  circuits, since the hysteresis properties also define the negative differential resistance of the memristive devices.
	
	We proposed a theoretical model which qualitatively describes the electric properties of the experimental devices. The model was able to predict a possibility for the temperature reset from a permanent LRS, later confirmed in experiment. Analysis of the dynamics of the proposed model  for varying device temperature $T_0$ allowed to reveal phenomena associated with change of residual time the system spends in LRS and HRS, which were also verified experimentally.
	
	Our results thus open a novel direction in efficient controlling devices and circuits involving the diffusive memristors, which is important for design and development of future neuromorphic computing and  AI systems. 
	
	\begin{acknowledgments}
		The authors would like to thank Sam Davies for the XPS analysis and acknowledge the use of the facilities within the Loughborough Materials Characterisation Centre.
		This work was supported by The Engineering and Physical Sciences Research Council (EPSRC), grant no. EP/S032843/1.
	\end{acknowledgments}

	\maketitle
		\FloatBarrier
	\bibliography{apssamp}
	
\end{document}